\documentclass[12pt,preprint]{aastex}

\usepackage{epsfig,graphicx,natbib,url,twoopt,epstopdf}
\usepackage[breaklinks]{hyperref}

\slugcomment{Accepted on 2015 September 23 for the publication of ApJ}

\begin{document}

\title{Origin of Interplanetary Dust \\through Optical Properties of Zodiacal Light}

\author{Hongu Yang\altaffilmark{1}, Masateru Ishiguro\altaffilmark{1}}

\affil{\altaffilmark{1}Department of Physics and Astronomy, Seoul National University, 599 Gwanak-ro, Gwanak-gu, Seoul 151-742, Republic of Korea}

\email{hongu@astro.snu.ac.kr}
\email{ishiguro@astro.snu.ac.kr}

\begin{abstract}
This study investigates the origin of interplanetary dust particles (IDPs) through the optical properties, albedo and spectral gradient, of zodiacal light. The optical properties were compared with those of potential parent bodies in the solar system, which include D--type (as analogue of cometary nuclei), C--type, S--type, X--type, and B--type asteroids. We applied Bayesian inference on the mixture model made from the distribution of these sources, and found that $>$90\% of the interplanetary dust particles originate from comets (or its spectral analogues, D--type asteroids). Although some classes of asteroids (C--type and X--type) may make a moderate contribution, ordinary chondrite--like particles from S--type asteroids occupy a negligible fraction of the interplanetary dust cloud complex. The overall optical properties of the zodiacal light were similar to those of chondritic porous IDPs, supporting the dominance of cometary particles in zodiacal cloud.
\end{abstract}


\section{Introduction}   \label{sec:introduction}
The purpose of this study is to investigate the origin of interplanetary dust particles (IDPs) taking into account the optical similarities and diversities between zodiacal light and reflections from minor bodies in the solar system, such as comets and asteroids. An enormous number of IDPs are distributed in interplanetary space. They are observable as scattered sunlight in the optical wavelength (zodiacal light) and as thermal radiation in the mid-- and far--infrared wavelengths (zodiacal emission). The IDPs cloud, occasionally referred to as a zodiacal cloud, erodes on a time scale of $10^{3}$--$10^{7}$ years (depending on the size and orbit, $<$ 1/100 of the age of the solar system) due to Poynting-Robertson drag, mutual collisions among the IDPs, and planetary perturbations \citep{2006A&ARv..13..159M, 1997ApJ...474..496G, 2001ESASP.495..609D}. The mass--loss rate around the Earth's orbit is estimated to be $\approx10^{3}$ kg s$^{-1}$ \citep{1985Icar...62..244G, 2005ApJ...621L..73M}. It is, therefore, natural to think that ongoing dust production, such as impacts or ice sublimation, is compensating for the erosion of the zodiacal cloud.

The origins of IDPs have been studied through the spatial distribution of the zodiacal light. Early research for explaining the spatial distribution expected large contribution from the asteroidal origin IDPs (see e.g. \citet{1996ASPC..104..143D}). Later, \citet{2002Icar..158..360H} compared the surface brightness distribution of zodiacal light, taken with the Clementine spacecraft onboard camera, to the inclination distributions of comets and asteroids and suggested that a significant fraction of dust particles at 1 AU are of cometary origin. \citet{2010ApJ...713..816N} further performed a numerical simulation for dust particles ejected from six different orbital groups (asteroid families, main belt asteroids, Jupiter family comets (JFCs), dormant JFCs, Halley-type comets, and Oort cloud comets), and compared the brightness distribution of the modeled zodiacal emission to that of zodiacal emission observed by an infrared space telescope. They suggested that $85\%$--$95\%$ of IDPs observable as zodiacal emission originate from JFCs.

Although these recent studies on the brightness distribution favor cometary sources, little is known about the origin of IDPs in terms of their optical properties. Recently, \citet{2013ApJ...767...75I} derived the geometric albedo of IDPs by comparing the brightness of the Gegenschein, a part of the zodiacal light enhanced by backward scattering enhancement, to the infrared model \citep{1998ApJ...508...44K}. The research provided possibilities to study the origin of IDPs from a different aspect than previous studies. In this study, we considered the origin of IDPs through the comparison of the albedo and spectral gradient of zodiacal light with those of the potential parent bodies, and present a discussion based on previous studies.

\section{Methodology}   \label{sec:Methodology}
The size distribution of IDPs was studied through lunar microcrater counting and \textit{in situ} flux measurements \citep{1985Icar...62..244G,1993JGR....9817029D}. These studies suggested that the effective cross section of IDPs around  the Earth's orbit is dominated by large (10--100 $\mu$m) particles. The opposition effect found in Gegenschein  supports the idea that IDPs, which make up the zodiacal light, are significantly larger than the optical wavelength \citep{2013ApJ...767...75I,2009Icar..203..124B}. Accordingly, we could assume that the optical properties of IDPs are similar to those of big objects, such as comets and asteroids. We thus postulated the albedo (A) and the spectral gradient ($S'$) of the IDPs according to those of potential dust sources in the following discussion.

\subsection{Albedo and spectral gradient of zodiacal light} \label{subsec:Spec}

The albedo of IDPs has been measured using several methods. \citet{1980Icar...43..373H} compared the zodiacal light brightness to the IDPs model derived from the Lunar microcrater records. \citet{1985Icar...62...54L} derived the albedo value of IDPs, which can explain the polarization distribution of zodiacal light. \citet{1988A&A...191..154D} compared the optical and infrared brightness at the solar elongation of 90\arcdeg. Recently, \citet{2013ApJ...767...75I} directly measured the geometric albedo of zodiacal light by comparing the optical and infrared \citep{1998ApJ...508...44K} flux at the anti--solar point. They deduced that the albedo of the smooth zodiacal light component is $A$=$0.06\pm0.01$, after subtracting weak fine--scale features associated with asteroidal collisional families, namely, asteroidal dust bands. In this paper, we adopted the albedo value in \citet{2013ApJ...767...75I}.

Small bodies in the solar system generally show linear spectra in a range around 4500--7500 \AA. It is useful to express the spectral index using the normalized reflectivity gradient, $S'$ [\%~10$^{-3}$ \AA]:
\begin{equation}
\label{eq:eq1}
S'=\frac{1}{\bar{S}}\frac{dS}{d\lambda},
\end{equation}
where $S$ is the reflectance, defined as the flux density of an object divided by the flux density of the Sun at the wavelength $\lambda$, and $\bar{S}$  and $dS/d\lambda$ denote the average reflectance and spectral gradient in the wavelength range, respectively. The spectra of zodiacal light have been measured mostly at infrared wavelengths from space \citep{1995Icar..115..199M, 1996PASJ...48L..47M, 2002ApJ...581..817F, 2003Icar..164..384R, 2009ASPC..418..395O, 2010ApJ...719..394T}. Since there is no spectrographic data available in the optical wavelength range, we derived the optical spectral gradient $S'$ by a log-linear fitting using compiled photometric data taken at different wavelengths around 4600~\AA \  \citep{1998A&AS..127....1L}. The regression formula for the ratio between the solar spectrum and the zodiacal light spectrum is given by,
\begin{equation}
\label{eq:eq2}
I_{\lambda} \propto \Big[ 1.0 + 0.9\times \log \left( \frac{\lambda}{5000~\mathrm{\AA} } \right) \Big] I_{\odot}
\end{equation}
where $I_{\lambda}$  and $I_{\odot}$ denote the flux densities of zodiacal light and the Sun at wavelength $\lambda$, respectively. Eq. (\ref{eq:eq2}) is applicable for zodiacal light in the spectral range of $\lambda\leq5000$~\AA \ at a solar elongation $>90\arcdeg$ \citep{1998A&AS..127....1L}. From Eq. (\ref{eq:eq2}), we obtained the spectral gradient of IDPs as $S'$=$8.5\pm1.0\ \% \cdot 1000~\mathrm{\AA}^{-1}$ at 4600~\AA. Note that we derived $S'$ at 4600~\AA\  in order to match the measured wavelength of the albedo \citep{2013ApJ...767...75I}.

\subsection{Data sources} \label{subsec:Data}

Turning now to the IDPs sources, we assumed that they originate from asteroids and comets. In addition, some IDPs may originate from interstellar space \citep{2002Icar..158..360H}. For asteroids, we considered five major taxonomic types, namely C--, X--, S--, B-- and D--types \citep{2013Icar..226..723D}, as input data. Since the optical properties of cometary nuclei are similar to those of D--type asteroids (one taxonomic type of asteroids), we do not discriminate D--type asteroids from cometary nuclei. We thus assumed that IDPs consist of dust particles from six populations: C--type, X--type, S--type, and B--type asteroids as representatives of asteroids, cometary nuclei (including D-type asteroids), and interstellar dust.

To create a template of the optical properties of six potential dust sources, we made use of catalogs of albedos and spectra of asteroids and comets. For asteroids, we used the Asteroid Catalog Using AKARI(AcuA) catalog as a dataset of albedos \citep{2011PASJ...63.1117U}, and the SMASSII catalog as datasets of spectral gradients \citep{bus1999study,2002Icar..158..106B,2004Icar..170..259B}. We found 274 C-type, 222 S-type, 191 X-type, 40 B-type and 33 D-type asteroids archived in both catalogs. For C-type, B-type and D-type asteroids, which show no obvious absorption, we used the spectral gradient values measured between $4350$ \AA \ and $9250$\AA \  \citep{2002Icar..158..106B,2004Icar..170..259B}. For the S-type and X-type asteroids, which may have an absorption band around $>7000~$\AA, we used the data at $4400~\mathrm{\AA} - 7000\mathrm{~\AA}$ \citep{bus1999study}. Albedos and spectral gradients of 10 cometary nuclei were compiled from various previous studies, shown in Table 1. For interstellar dust, we used the optical properties of average galactic dust particles at 4600~\AA, that is, $A$=0.67 and $S'$=$-23\pm1 \ \% \cdot 1000~\mathrm{\AA}^{-1} $ \citep{2003ApJ...598.1017D}. We ignored some taxonomic types of asteroids, such as K--type, L--type and V--type as discussed in Section 4.

\begin{deluxetable}{cccccc}
\tablecaption{Optical properties of cometary nuclei.}
\tablewidth{0pt}
\tablehead{\colhead{Name} & \colhead{type} & \colhead{albedo} & \colhead{spectral gradient} & \colhead{references}}
\startdata
1P/Halley & Halley type & 0.043 & 7.5 & a, c, g \\
2P/Encke & Encke type & 0.050 & 11 & h, j, r \\
9P/Tempel 1 & Jupiter family & 0.056 & 12.5 & q \\
10P/Tempel 2 & Jupiter family & 0.022 & 20 & e, f \\
28P/Neujmin 1 & Jupiter family & 0.025 & 11.8 & b, k, p \\
49P/Arend-Rigaux & Jupiter family & 0.028 & 10.4 & d, i \\
67P/Churyumov-Gerasimenko & Jupiter family & 0.047 & 10 & o, s, t, u \\
103P/Hartley 2 & Jupiter family & 0.045 & 8.1 & v \\
162P/Siding Spring & Jupiter family & 0.034 & 9.2 & m, n \\
C/2001 OG$_{108}$ (LONEOS) & Halley type & 0.040 & 9 & l \\
\enddata
\tablecomments{a - \citet{1986Natur.321..259S}, b - \citet{1987ApJ...316..847C}, c - \citet{1987A&A...187..807K}, d - \citet{1988ApJ...324.1194M}, e - \citet{1989ApJ...347.1155A}, f - \citet{1989AJ.....97.1766J}, g - \citet{1989A&A...213..487T}, h - \citet{1990Icar...86...69L}, i - \citet{1993Icar..104..138L}, j - \citet{2000Icar..147..145F}, k - \citet{2002EM&P...89..117C}, l - \citet{2005Icar..179..174A}, m - \citet{2006AJ....132.1346C}, n - \citet{2006AJ....132.1354F}, o - \citet{2006A&A...458..669L}, p - \citet{2007AJ....134.1626C}, q - \citet{2007Icar..187...41L}, r - \citet{2008A&A...489.1337B}, s - \citet{2008A&A...489..777L}, t - \citet{2008A&A...490..377T}, u - \citet{2009AJ....137.4633K}, v - \citet{2013Icar..222..559L} }
\end{deluxetable}

Figure 1 shows the relationship between the albedos and spectral gradients for the IDPs and the potential parent bodies described above. In the  diagram, the datum of IDPs is located within the population of comets and its spectral analog, D-type asteroids, suggesting that the major constituents are of cometary origin. In the following section, we will further investigate the contribution of each population using statistical analysis.

\begin{figure}
\figurenum{1}
\epsscale{0.8}
\plotone{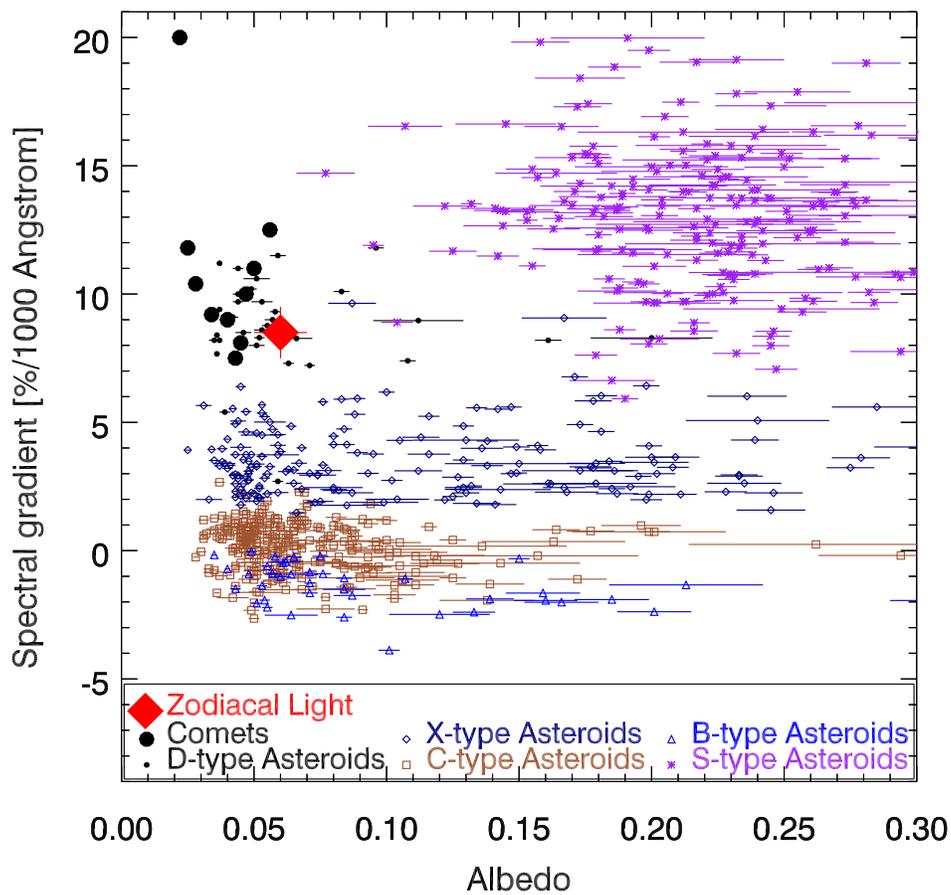}
\caption{Spectral gradients $S'$ with respect to albedos $A$ of asteroids, comets and zodiacal light. Uncertainties of albedos are appended in the plot. The 1$\sigma$ measurement uncertainties in spectral gradients are ordinarily about 0.7 $\% \cdot 1000 \mathrm{~\AA}^{-1}$.}
\end{figure}

\subsection{Bayesian analysis}

When we chose objects from a type of population and calculated the correlation coefficients between the albedos and spectral gradients, the absolute values were as low as -0.14, -0.04, 0.16, -0.36 and -0.11, for the C--type, S--type, X--type, B--type and D--type asteroids, respectively. Therefore, we considered these two properties, the albedo and spectral gradients, as being independent of each other. Within a population, we simply assumed that the albedo follows a log-normal distribution whereas the spectral gradient has a Gaussian distribution. We fit the distributions shown in Figure 1 to the model distributions shown in Figures 2 and 3. Because interstellar dust particles, which have limited contribution to IDPs, have optical properties that are very different from those of solar system objects, we assumed that the interstellar dust particles have a fixed albedo value of $A=0.673$ and a spectral gradient of $S'=-23.2\pm 0.8 \ \% \cdot 1000\mathrm{~\AA}^{-1} $, and did not consider its statistical distributions in the following analysis. 

\begin{figure}
\figurenum{2}
\epsscale{0.8}
 \plotone{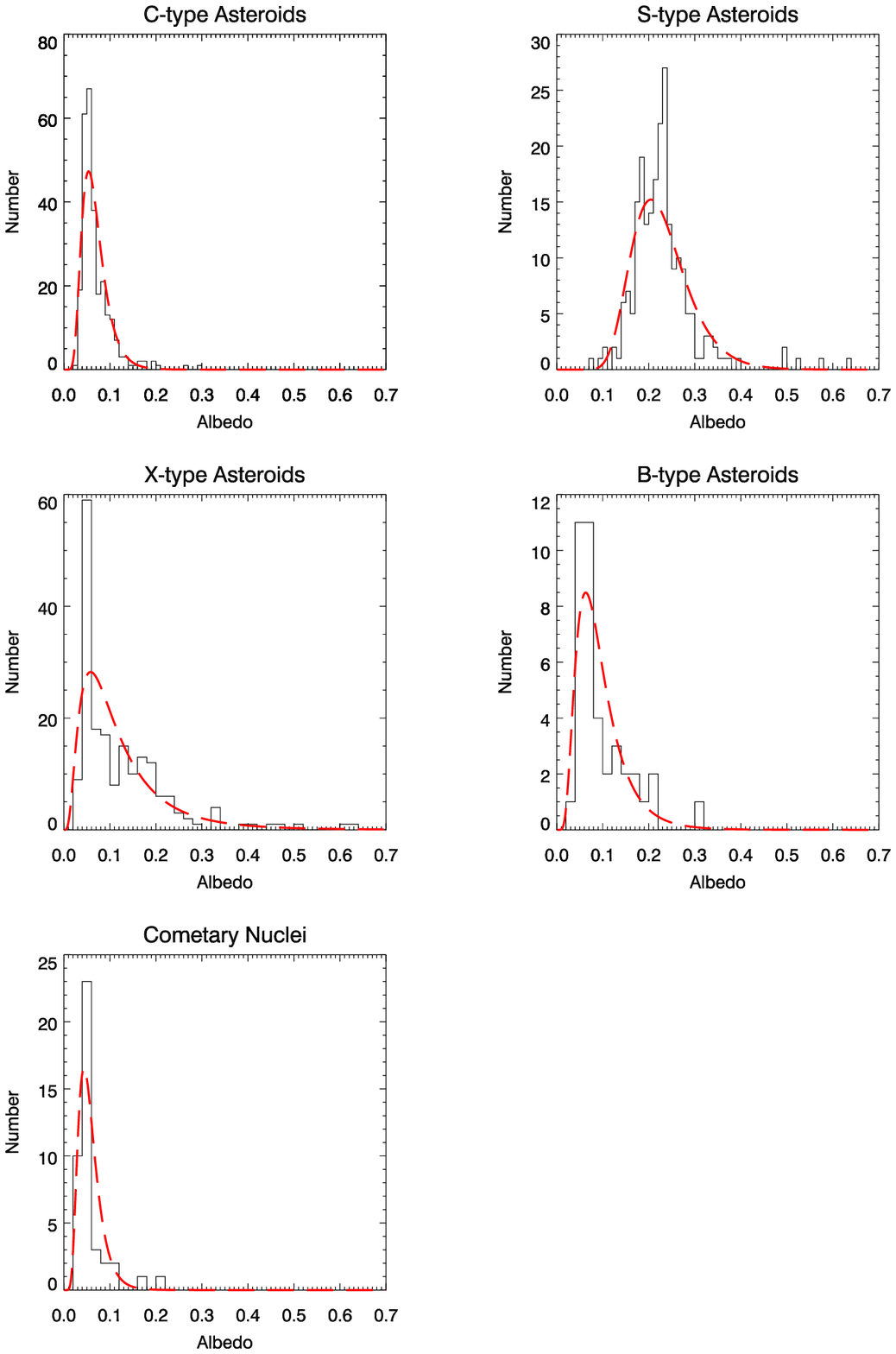}
\caption{Albedo distribution of C--type, S--type, X--type, B--type, cometary nuclei (including D--type asteroids) \citep{2011PASJ...63.1117U}. Black solid lines are histograms for the given types. Red dashed lines are the lognormal distributions calculated from the mean and standard deviation of logarithms.}
\end{figure}

\begin{figure}
\figurenum{3}
\epsscale{0.8}
\plotone{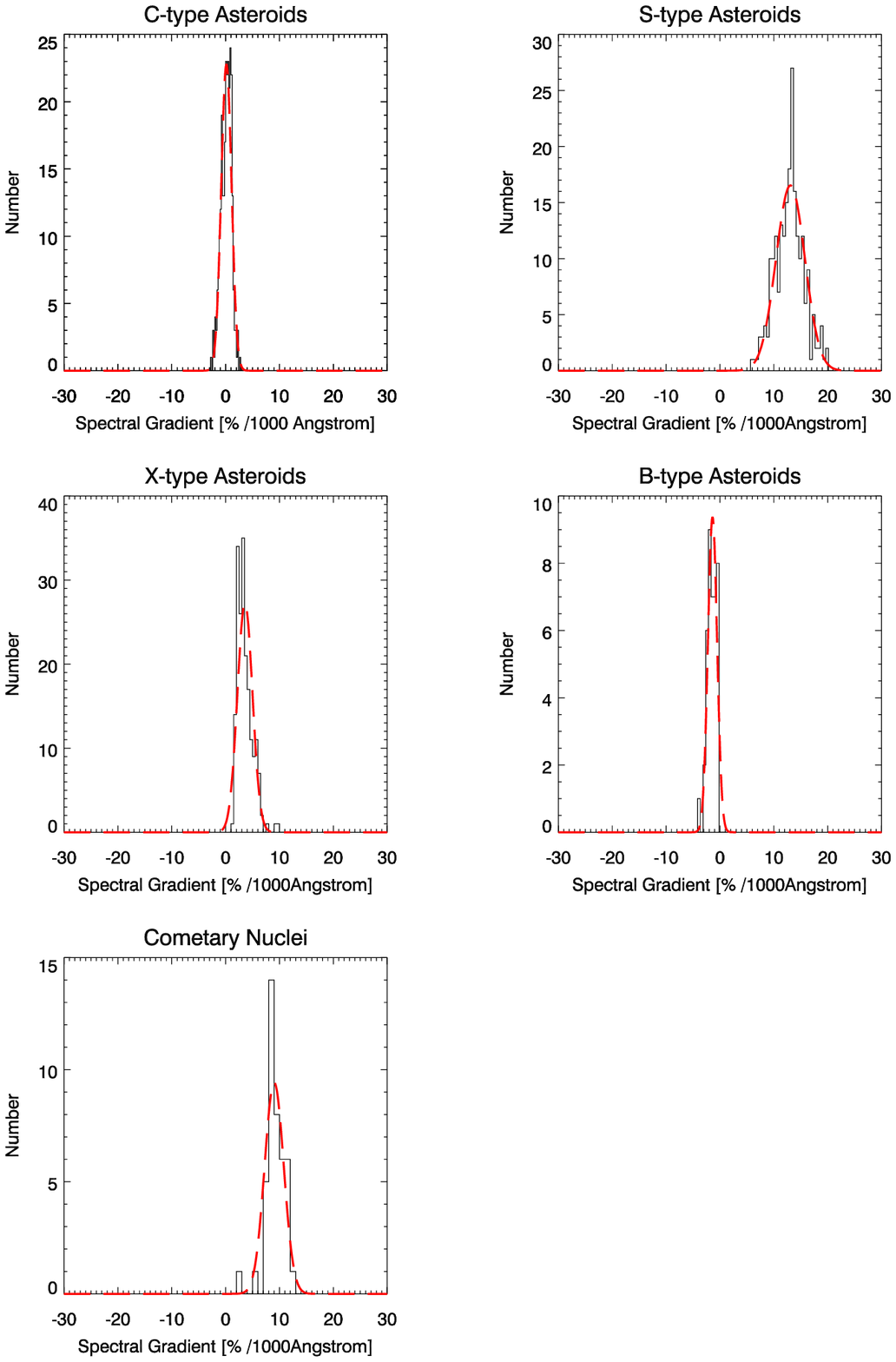}
\caption{Spectral gradient distribution of C--type, S--type, X--type, B--type, cometary nuclei (including D--type asteroids) \citep{bus1999study,2002Icar..158..106B}. Black solid lines are histograms for the given types. Red dashed lines are the Gaussian distributions calculated from the mean and standard deviation.}
\end{figure}

We made a 2\% grid for the possible combinations of fractional contributions from the source populations. Then, through linearly combining the probability distributions of the source populations according to the given contribution, we could generate a mixture probability distribution for an optical property of a single dust particle. Different populations were weighted according to the average albedos. From the probability distribution for a single particle, we calculated the expected average values of both albedo and spectral gradient for the IDPs complex using Monte Carlo (MC) simulations. At the every grid point, a MC simulation with 500 sample particles was generated 3000 times. Under these conditions, the expected average values follow a Gaussian distribution, with a standard deviation of less than 10\% of the uncertainty on the zodiacal light measurement. The probability of obtaining the measured optical properties was calculated at each grid point. By applying Bayesian inference with a flat prior, the probability was accepted as that of the assumed contribution of the grid point represent real situation.

\section{Results}   \label{sec:Results}

Table 2 shows the resulting contributions from the individual sources to the IDPs cloud. To derive the ranges (which are shown as plus and minus signs in the Table 2), we made contours with the same probability in the six dimensional grid, calculated the total probability within contours around the most probable case, and derived the range with the 68.3 \% confidence interval. We found that cometary nuclei (including D--type asteroids) are the primary contributors ($\sim 94$\%) to the IDPs cloud as predicted in the Section 2.2. The remaining part ($\sim 6$\%) is originated from the C--type, X--type and B--type asteroids. S--type asteroids and interstellar dust have an insignificant contribution to the IDPs ($\sim 0$\%). Figure 4 shows the marginalized probability distributions of the four major populations.

\begin{deluxetable}{cc}
\tablecolumns{2}
\tablewidth{0pc}
\tablecaption{Contribution of 6 source populations to the IDPs}
\tablehead{\colhead{Population} & \colhead{contribution}}
\startdata
Cometary nuclei & 94$^{+6}_{-26}$\% \\
B-type asteroids & 4$^{+18}_{-4}$\% \\
X-type asteroids & 2$^{+24}_{-2}$\% \\
C-type asteroids & 0$^{+26}_{-0}$\% \\
S-type asteroids & 0$^{+10}_{-0}$\% \\
Interstellar dust & 0$^{+2}_{-0}$\% \\
\enddata
\end{deluxetable}

\begin{figure}
\figurenum{4}
\epsscale{1}
\plotone{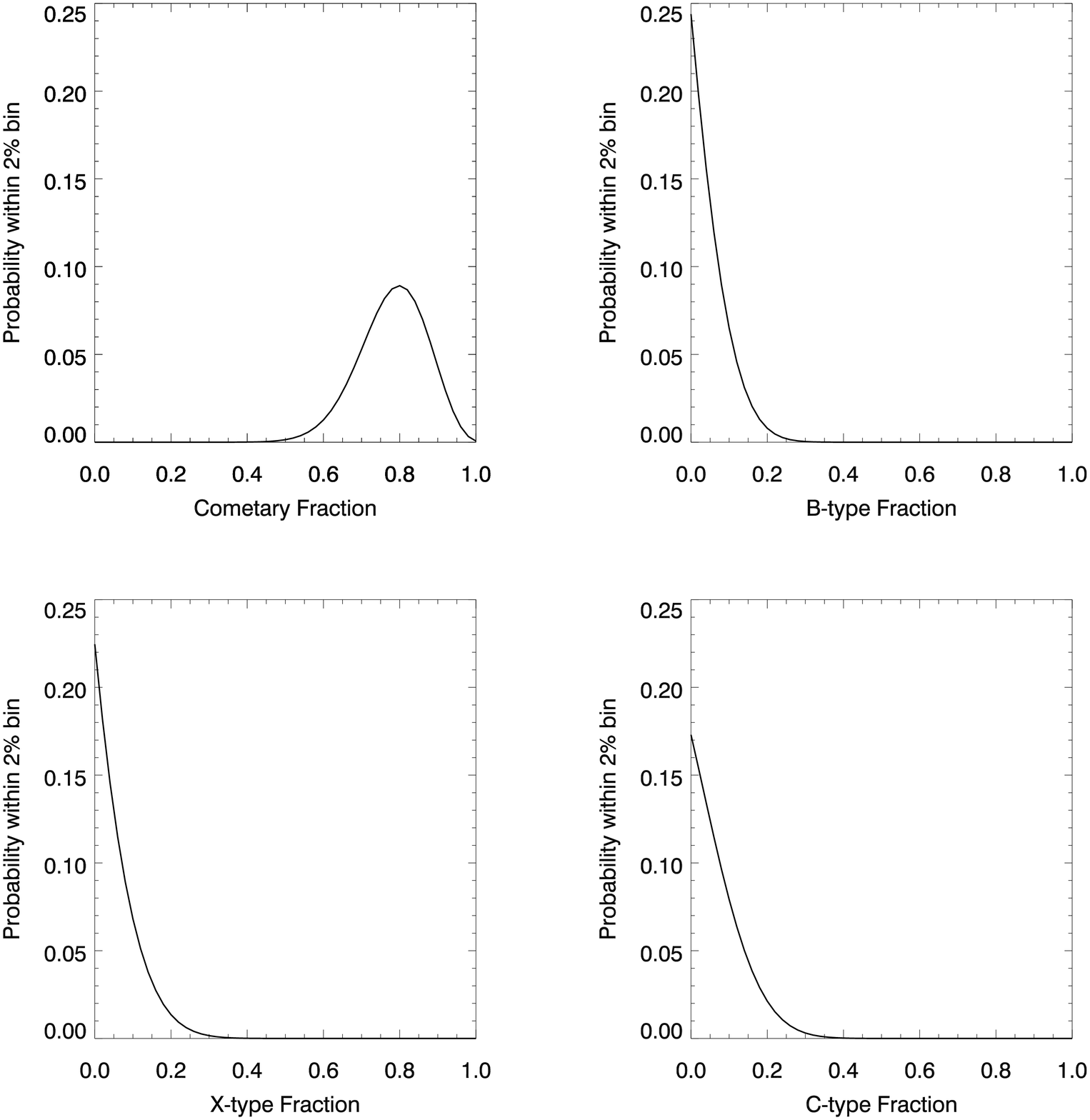}
\caption{(a) - (d) Marginalized probability for the fractions about cometary (D--type asteroids), B--type, X--type and C--type asteroids. The probability for the vertical axis is the values integrated over 2\% bin.}
\end{figure}

\section{Discussion}   \label{sec:Discussion}

\subsection{Feasibility of the method}   \label{sec:Feasibility of the method}

To assess the feasibility of our approaches described above, let us discuss the three following points.

Firstly, we should consider the validity of the source populations. There are a wide variety of objects in the solar system, however, only six types of sources (five types of asteroids, comets and interstellar dust particles) are considered in this paper. Recently, the mass fraction from different taxonomic types of asteroids was studied using new multi-filter photometric survey data. \citet{2013Icar..226..723D} suggested that C--type asteroids account for more than 50\% of the mass in the main belt. Although S--types, P--types, B--types, and V--types have moderate fractions ($\sim 10\%$ each of the total mass of asteroids in the main belt), other asteroids such as K--types, L--types and A--types only have minor contributions of $<1\%$. P--types are included in X--types in our assumption. Thus, we considered all but one, namely V--types, major asteroids in this paper. We conjecture that V--types cannot contribute to the IDPs cloud because they have very large albedos ($A=0.30$, \citet{2011PASJ...63.1117U}). In addition, the mass fraction of V--types is very small (0.01\%) when we exclude the largest objects in the taxonomic type (i.e. (4) Vesta). Meanwhile, the photopolarimeters on the Pioneer 10 and 11 spacecrafts revealed that the zodiacal light brightness is negligible beyond 3.3 AU \citep{toller1981study}. Those observations suggest that the contribution from outer objects such as the Kuiper--belt objects (KBOs) may not be as large as those from asteroids, when we think about the previous dynamic studies pointing out that the dust particles from KBOs have peak densities outside of the Jovian orbits \citep{2012GeoRL..3915104P,2014AJ....147..154V}. The optical properties of the Centaurs show bimodality. Inactive Centaurs show ultrared spectra similar to KBOs, and active Centaurs have colors and albedos similar to cometary nuclei \citep{2008ssbn.book..161S, 2012A&A...539A.144M}. In this paper, inactive ultrared Centaurs were ignored along with KBOs and active Centaurs were treated as cometary nuclei. Some cometary nuclei have optical properties that are different from those of D--type asteroids, as in the cases of 95P/Chiron and 107P/Wilson-Harrington. We did not include these kinds of objects in this study. We do not know how many such objects exist, but these non D--type asteroidal nuclei are similar to other kinds of asteroidal groups in terms of optical properties. Therefore each population of an asteroidal group should be understood to include possible cometary nuclei whose optical properties are similar to the group. If we subdivide the X--type asteroids into E--type, M--type and P--type, the results remains same, with only the confidence interval become worsen, because the optical properties of P--type asteroids are similar to those of D--type asteroids.

Secondly, we should consider the time--evolution of the optical properties via space weathering. We assumed that the optical properties of dust particles resemble those of the source objects. However, it may not be true in some populations. Since the Poynting--Robertson lifetime of silicaceous dust particles of 1 mm size is about $2\times10^{7}$ years when released into circular orbit from 2.5AU \citep{2006A&ARv..13..159M} while the time scale of the space weathering is more than an order of magnitude shorter than the lifetime \citep[$\sim 7 \times 10^{5}$ years for S--type asteroids.]{2013Icar..225..781S}, it is reasonable to assume that surfaces on both silicaceous IDPs and S--type asteroids are altered by space weathering and therefore have similar optical properties. However, the space weathering of C--type, X--type and B--type asteroids are not well known, although there are studies, e.g. \citet{2004Icar..170..214M}. Therefore, we cannot clearly discuss the optical surface maturation of IDPs originated from these asteroids. Furthermore, cometary dust particles remain within the interplanetary space longer than the active lifespan of cometary nuclei \citep[$\sim$12,000 years for the ecliptic comets]{1997Icar..127...13L}, therefore the relation between the optical properties of cometary dust particles and the surfaces of active cometary nuclei is not direct. If we regard the cometary nuclei and D--type asteroids as identical, there are studies that imply that the spectra of D--type asteroids would not change significantly over time. D--type asteroids were found in the inner main belt \citep{2014Icar..229..392D}, and Phobos, possibly captured D--type asteroids, have optical properties of a D--type asteroid after remaining in the inner solar system for billions of years \citep{2013ApJ...777..127P,2014MSAIS..26...67P}. Even though these objects have albedo values slightly higher than the average of D--type asteroids, their albedos and spectral gradients are still in the range of D--type asteroids. In other direction, according to the laboratory experiments on the Targish Lake meteorites, which have spectra similar to D--type asteroids, the continuum spectrum changed to the bluer direction after being exposed to laser radiation \citep{2012LPICo1667.6109H}. If these results can be applied in our case, the contribution of cometary nuclei would increase. If we think about the even now dominant cometary contribution, we can conclude that this assumption does not alter the conclusion of this paper.

Thirdly, we should consider the effects of simplification in this study. We assumed that the optical properties are randomly dispersed within a population, however this may not be true. As shown in \citet{2013ApJ...762...56U}, there is a relation between orbital elements and optical properties. We want to emphasize that the differences in the optical properties between different types of sources are an order of magnitude bigger than the differences between sub-groups of different types of sources, as shown in Figure 1. Furthermore, we ignored the weak correlation between the albedo and spectral gradient. The relation was non-negligible for B--type asteroids. There is a possibility that the interstellar dust, which entered the solar system, has a different composition compared to the average dust particles in the Milky Way galaxy \citep{2010ARA&A..48..173M}, but we ignored this possibility. However, we want to justify these simplifications because the contributions from B--type asteroids and interstellar dust are almost negligible. The SMASSII catalogue is not bias-free \citep{2003Icar..162...10M}, and we do not know the optical properties of unbiased populations, but we ignored the effect from bias. We hope that the large measurement uncertainties in the optical properties of zodiacal light cover the consequences of the bias.

\subsection{Comparison with IDP samples}

IDPs are nowadays collected in the Antarctic ice or the stratosphere around 20--25 km altitude using aircrafts, and they are well studied through laboratory investigations \citep{1985AREPS..13..147B,1998M&PS...33..565E}. Because such particles should contribute to the zodiacal light before they arrive on Earth, it is important to compare our result to IDPs samples. It is known that there are two major IDPs groups, which are referred to as "chondritic smooth" (CS) and "chondritic porous" (CP). CS IDPs are composed of low porosity materials, predominantly hydrated layer silicates \citep{1985ApJ...291..838S}. CP IDPs have large porosities of about $\sim$70\%. CP IDPs are dominated by anhydrous minerals. It is likely that CP IDPs originate from comets on the ground of mineralogical and petrographical properties \citep{2003TrGeo...1..689B}. When the Earth passed through the dust stream of 26P/Grigg-Skjellerup (one of the JFCs), it was expected that 1--50\% of the total collected dust larger than 40 \micron \ could originate from that comet \citep{2002M&PS...37.1491M}. These were actually CP type IDPs with primitive anhydrous composition, supporting the assumption of that CP IDPs are of cometary origin \citep{2009E&PSL.288...44B}. CS IDPs are thought to be derived from primitive (not differentiated) asteroids, because comets do not exhibit spectral signatures of hydrated silicate while asteroids do \citep{2015Icar..245..320M}.

\citet{1996M&PS...31..394B} measured the reflectance spectra of IDPs samples in the optical wavelength and found that CS IDPs generally exhibit flat spectra with weak fall-off from 6000 \AA \ towards 8000 \AA \ (similar to CI and CM meteorites or C--type asteroids with $S'\sim$ 0) whereas CP IDPs exhibit upward spectra ($S'>$ 0) without remarkable curvature, although these IDP samples have a variety of albedo values and spectral slopes. Figure 5 compares the synthesized spectrum of zodiacal light based on our mixing model of small bodies with those of most typical CS IDPs (W7040A15) and CP IDPs (W030A5) \citep{1996M&PS...31..394B}. We also show the input spectrum in Figure 5, which was obtained from a multi-band photometry in \citet{1998A&AS..127....1L} and anti-solar point observation \citep{2013ApJ...767...75I}. The synthesized spectrum of zodiacal light is similar to that of CP IDPs (W030A5) in that it shows a low albedo value and positive slope ($S'>$ 0) but it is different from that of CS IDPs in that it does not show a positive slope beyond $\sim 6500 \AA$. It should be noted that the fall-offs below 4500 \AA \ in IDPs signals are artifacts of the measurements caused by small size effects and should be ignored for the comparison \citep{1996M&PS...31..394B}. The spectral similarity leads to our assumption that the interplanetary dust complex is dominated by CP IDPs (i.e. dominance of cometary particles in zodiacal cloud). It is also curious to notice that CP IDPs tend to have cluster structures of 20-100 \micron \ \citep{1996M&PS...31..394B}. The size is in accordance with the effective size of zodiacal light dust particles evaluated by the IDPs size distribution model \citep{1985Icar...62..244G} .

\begin{figure}
\figurenum{5}
\epsscale{1.0}
\plotone{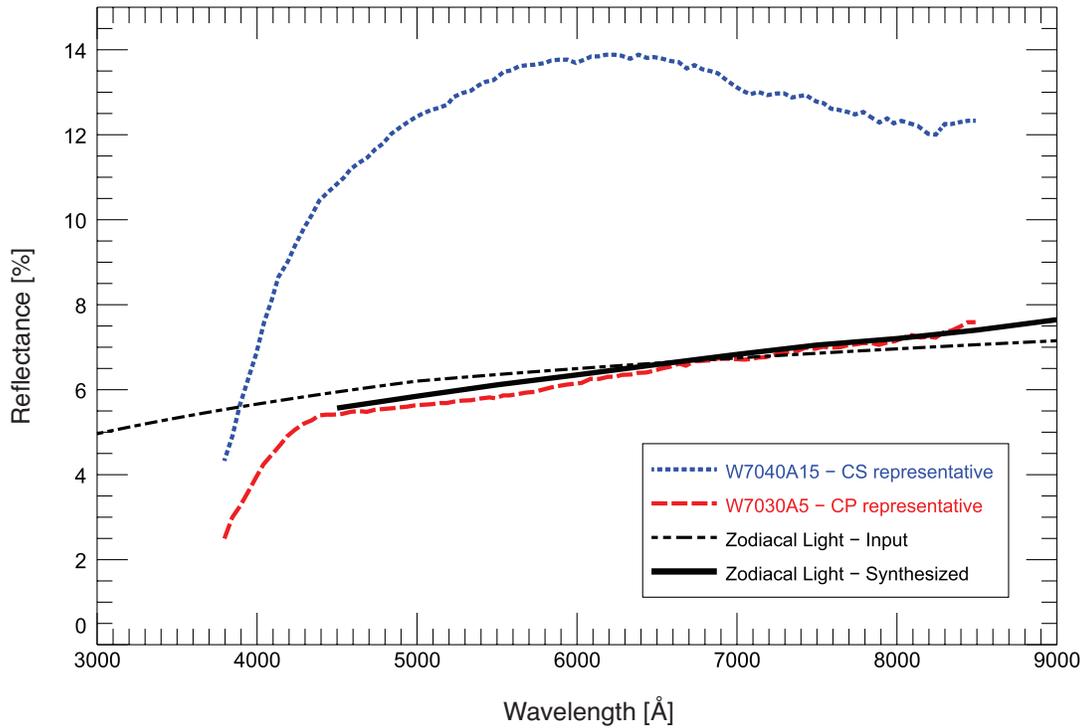}
\caption{Comparisons of spectra between our synthesized zodiacal light model (thick continuous line), CS and CP IDPs (dotted and dashed lines) \citep{1996M&PS...31..394B}. The observed reference spectrum of zodiacal light is also shown (see Section 2.1). Note the drop-offs in the IDP spectra at less than 4000 \AA \ artifacts.}
\end{figure}

There seems to be a difference on the quantitative estimates of IDPs origins between laboratory investigations of IDPs samples and ours.
\citet{2003TrGeo...1..689B} studied 200 chondritic IDPs from the stratosphere and found that about a half of them are classified into CP IDPs.
Similarly, \citet{2015E&PSL.410....1N} investigated micrometeorite samples  in Antarctica and suggested
$\sim$25\% or even less are categorized into CP IDPs. Whereas we acknowledge that these laboratory investigations 
provide reliable results regarding the fraction of CS and CP IDPs fallen to Earth, we would draw attention to the sampling
bias of the laboratory studies of IDPs. Asteroidal dust particles could be collected on Earth more selectively than cometary dust particles because of the orbital properties. The impact cross-section of asteroidal dust particles to the Earth is a few times larger than that of cometary ones. Furthermore, the impact cross--section of the Earth can be a few 1000 times larger for dust particles trapped in quasi-satellite resonance, which favors asteroidal particles \citep{2013Icar..226.1550K}.

\subsection{Comparison with previous studies}

Our results are in agreement with kinematic, dynamical studies based on spatial distribution of zodiacal light. Numerical simulations of \citet{2010ApJ...713..816N} concluded that $\gtrsim 90$\% of the zodiacal emission comes from JFC originated IDPs, and $\lesssim 10$\% from the Oort cloud comets or asteroids. These results coincide with our results of $>90$\% cometary contribution. However, we could not optically distinguish different sub-populations of comets in this study. Therefore, we cannot know how large of a fraction of cometary IDPs originated from JFCs.

We compared our results with infrared spectroscopic observations of zodiacal light. By using mean albedo values and contributions from this study and adopting a typical visual-infrared spectrum of the populations from \citet{2002Icar..158..146B}, we synthesized the model spectrum of zodiacal light, and extrapolated it to the near-infrared wavelength, shown in Figure 6. At wavelengths shorter than 1.6 \micron, the observed NIR spectra of the zodiacal light \citep{2010ApJ...719..394T} are similar to the synthesized spectra in large-scale, but have absorption-like dark wavelengths around 1.3--1.4 \micron. In this wavelength regime, the spectra of average B--type, C--type and X--type asteroids are bluer than those of zodiacal light, as expected in this study. The observed spectra have large uncertainties at wavelengths longer than 1.6 \micron. It is, therefore, hard to conclude clearly, but the observed spectra are closer to those of D--type, C--type or X--type asteroids than to those of S--type or B--type asteroids. Overall, the synthesized spectrum in this work is consistent with the observed spectrum from rocket-borne observations \citep{1995Icar..115..199M, 2010ApJ...719..394T} and recent space observations of AKARI \citep{2013PASJ...65..119T}, but it does not agree with the space observation of IRTS \citep{1996PASJ...48L..47M}. The observed NIR spectra of zodiacal light are similar to those of D--type asteroids or cometary nuclei, which is in agreement with this work. Keeping in mind that we applied extrapolation through the model from optical observation, this match in the NIR reflected spectrum supports our conclusion of dominating cometary contribution.

\begin{figure}
\figurenum{6}
\epsscale{1.0}
\plotone{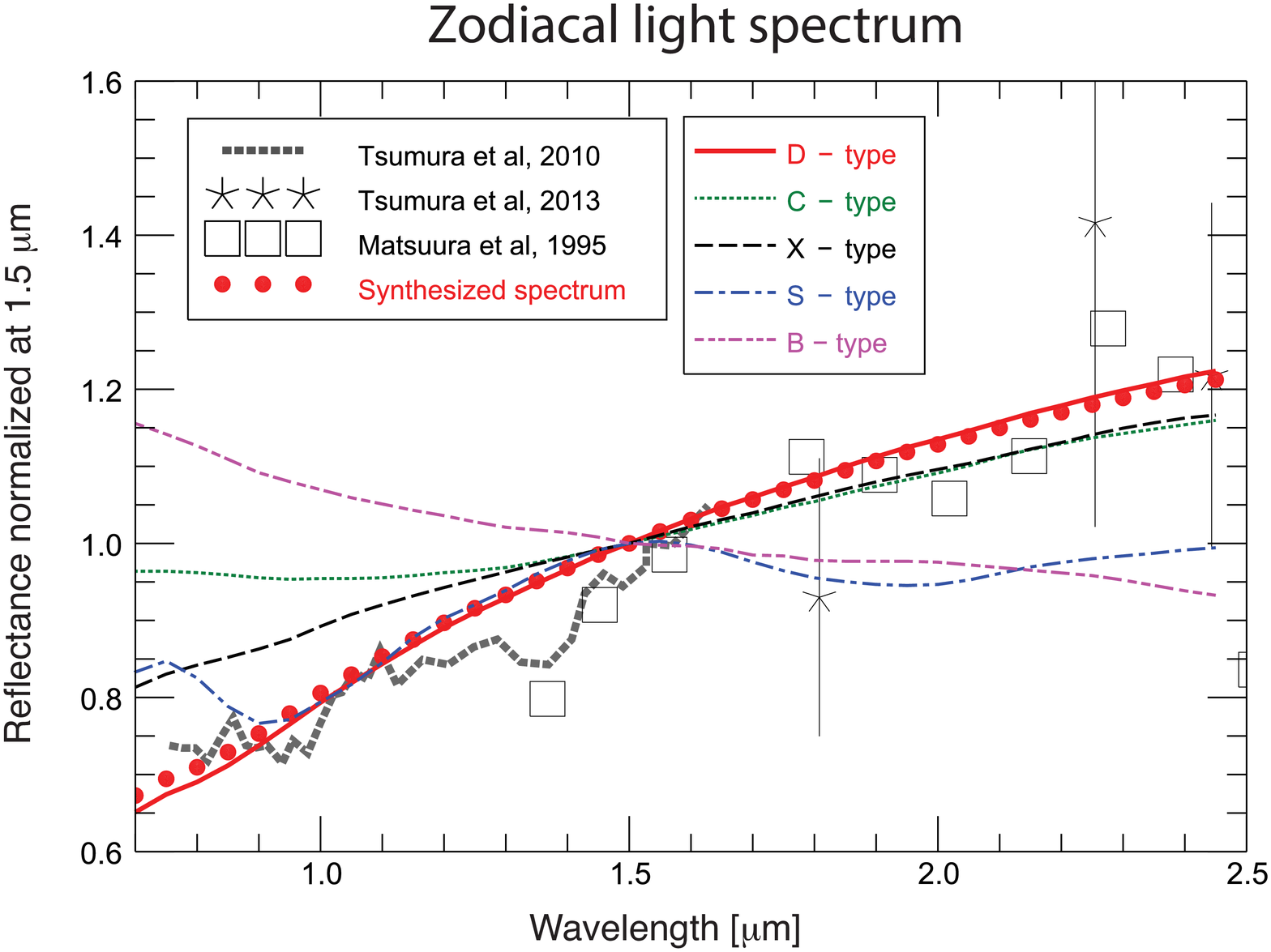}
\caption{Comparison of near-infrared zodiacal light spectra between our synthesized model and observed data \citep{1995Icar..115..199M,2010ApJ...719..394T,2013PASJ...65..119T}. The S-220-11 rocket data at an ecliptic latitude of 10$^{\circ}$ from \citet{1995Icar..115..199M} was used in the figure. The template spectra of each type of asteroid are from \citet{2002Icar..158..146B}, and the solar spectra are from \citet{2004SolEn..76..423G}. The templates, synthesized spectra, and the data from \citet{2010ApJ...719..394T} are normalized at 1.5 $\mu$m, and the data from \citep{1995Icar..115..199M,2013PASJ...65..119T} are scaled to match our model spectrum at 1.8--2.5 \micron.}
\end{figure}

\acknowledgments We wish to acknowledge Dr. Fumihiko Usui for his valuable comments to the manuscript, and Dr. Kohji Tsumura for the kind offering of his data. This work was supported by the National Research Foundation of Korea (NRF) funded by the South Korean government (MEST) (Grant No. 2012R1A4A1028713).

\bibliographystyle{apj}
\bibliography{Draft_150901}

\end{document}